\documentclass[namedreferences]{SSRv}
\usepackage{graphicx}
\usepackage{times} 
\usepackage{url}
\def\ebv{\hbox{$E(B-V)$}}

\def\Bis{\hbox{$B_{\rm IS }$}}
\def\Bpar{\hbox{$B_\mathrm{|| }$}}
\def\Bperp{\hbox{$B_{\perp }$}}

\def\glong{\hbox{$\ell$}}
\def\glat{\hbox{$b$}}
\def\micron{\hbox{$\mu$m}}
\def\mG{\hbox{$\mu$G}}
\def\cmtwo{\hbox{cm$^{-2}$}}
\def\cc{\hbox{cm$^{-3}$}}
\def\kms{\hbox{km s$^{-1}$}}
\def\HeI{\hbox{He$^\circ$}}
\def\NaI{\hbox{Na$^\circ$}}
\def\MgI{\hbox{Mg$^\circ$}}
\def\MgII{\hbox{Mg$^+$}}
\def\NH{\hbox{$N$(H)}}

\def\nH{\hbox{$n$(H)}}
\def\nel{\hbox{$n$(e)}}
\def\nHI{\hbox{$n$(H$^\circ$)}}
\def\nHeI{\hbox{$n$(He$^\circ$)}}
\def\nHmean{\hbox{$< n$(H)$>$}}
\def\NHH2{\hbox{$N$(${\rm H^{ \rm o } + 2H_2}$)}}
\def\HH{\hbox{H$_2$}}

\def\HII{\hbox{H$^{ \rm + }$}}
\def\NII{\hbox{N$^{ \rm + }$}}
\def\NI{\hbox{N$^\circ$}}
\def\HI{\hbox{H$^{ \rm o }$}}
\def\NHI{\hbox{$N$(H$^{\rm o }$)}}

\def\NHH{\hbox{$N$(H$_{\rm 2 }$)}}

\newcommand\apj{{ApJ}}
\newcommand\aap{{A\&A}}
\newcommand\aj{{AJ}}
\newcommand\mnras{{MNRAS}}
\newcommand\ssr{{Space~Sci.~Rev.}}
\newcommand\apjl{{ApJ}}
\newcommand\apjs{{ApJS}}

\begin{document}
\begin{article}
\begin{opening}

\title{The Local Bubble and Interstellar Material Near the Sun}
\runningtitle{Local Bubble and ISM Near the Sun}

\author{P. C.  \surname{Frisch}$^{1,*}$}
\institute{
$^1$Dept. of Astronomy and Astrophysics, University of Chicago, 5640 S. Ellis Ave., Chicago, IL 60637, USA \\
($^*$Author for correspondence, Email: frisch@uchicago.edu)}

\runningauthor{P. C. Frisch} 

\received{12 February 2007} \revised{13 February 2007} \accepted{14 February
2007}

\begin{abstract}

The properties of interstellar matter at the Sun are regulated by our
location with respect to a void in the local matter distribution,
known as the Local Bubble.  The Local Bubble (LB) is bounded by
associations of massive stars and fossil supernovae that have
disrupted dense interstellar matter (ISM), driving low density
intermediate velocity ISM into the void.  The Sun appears to be
located in one of these flows of low density material.  This nearby
interstellar matter, dubbed the Local Fluff, has a bulk velocity of
$\sim 19$ \kms\ in the local standard of rest.  The flow is coming
from the direction of the gas and dust ring formed where the Loop I
supernova remnant merges into the LB.  Optical polarization data
suggest the local interstellar magnetic field lines are draped over
the heliosphere.  A longstanding discrepancy between the high thermal
pressure of plasma filling the LB and low thermal pressures in the
embedded Local Fluff cloudlets is partially mitigated when the ram
pressure component parallel to the cloudlet flow direction is
included.  

\end{abstract}

\keywords{ISM: general, ISM: abundances}

\end{opening}
 
\section{Introduction} \label{sec:intro}

The existence of an area clear of interstellar material around the
 Sun, now known as the Local Bubble, was discovered as an underdense
 region in measurements of starlight reddening (Fitzgerald, 1968).
This underdense region is traced by color excess measurements showing
\ebv$< 0.01$
 mag,\footnote{\ebv=$A_\mathrm{B} - A_\mathrm{V}$, where
 $A_\mathrm{B,V}$ is the attenuation in units of magnitude in the blue
 (B) and visible (V) bands, respectively.} and extends beyond
100 pc in the galactic longitude interval $\ell=180^\circ - 270^\circ$.  
In the plane of Gould's Belt, the
Local Bubble boundaries (``walls'') are defined by interstellar material
 (ISM) associated with star forming regions.  At high galactic
 latitudes the Local Bubble boundaries are defined by interstellar gas
 and dust displaced by stellar evolution, particularly supernova in
 the Scorpius-Centaurus Association.  Supernovae exploding into
 pre-existing cavities created by massive star winds displace ISM and
 the interstellar magnetic field into giant magnetized bubbles
 hundreds of parsecs in extent. The location of the Sun within such a
 void regulates the interstellar radiation field at the Sun,
 and the composition and properties of the ISM surrounding the
 heliosphere.  

The Local Interstellar Cloud (LIC), defined by the velocity of interstellar 
\HeI\ inside of the heliosphere, is one cloudlet in a low density ISM flow 
known as the Local Fluff.  The Local Fluff has an upwind direction towards 
Loop I and the Scorpius-Centaurus Association (SCA).  This flow, with a
best-fit local standard of rest (LSR) velocity of $\approx 19.4$
 \kms, appears to be a break-away fragment of the Loop I superbubble
 shell surrounding the SCA, which has expanded into the low density
 interior of the Local Bubble (\S \ref{sec:origin}, Frisch, 1981;
 Frisch, 1995; Breitschwerdt et al., 2000).

This paper is in honor of Prof. Johannes Geiss, founder the
 International Space Sciences Institute (ISSI).  Many of the
 contemporary space topics discussed at ISSI meetings, such as the
 heliosphere, the Local Interstellar Cloud, cosmic ray acceleration
 and propagation, and the composition of matter, are influenced by the
 solar location inside of the Local Bubble.

\section{Origin and Boundaries of the Local Bubble}

\subsection{Origin} \label{sec:origin}
The Local Bubble void has been created by star formation processes
that have occurred during the past 25--60 Myrs in the corotating region
of the Milky Way Galaxy near the solar location of today.  About
25--60 Myrs ago a blast wave evacuated a low density region at the
present location of the Sun, and compressed surrounding molecular
clouds to initiate the formation sequence of the massive OB stars now
attributed to Gould's Belt.  Gould's Belt denotes the system of
kinematically related massive OB stars within $\sim 500$ pc of the
Sun, which form a localized plane tilted by $\sim 18^\circ$ with
respect to the galactic plane.  The center of Gould's Belt is 104 pc
from the Sun towards \glong=180$^\circ$, and with an ascending node
longitude of 296$^\circ$ (Grenier, 2004).  The
Sun is moving away from the center of Gould's Belt, and is closest to
the Scorpius-Centaurus rim.  
Overlapping superbubbles shape the Local Bubble void 
(Frisch, 1995; Heiles, 1998; Maiz-Appellaniz, 2001). 

\begin{figure}
\centerline{\includegraphics[angle=90,clip=,width=1.0\textwidth]{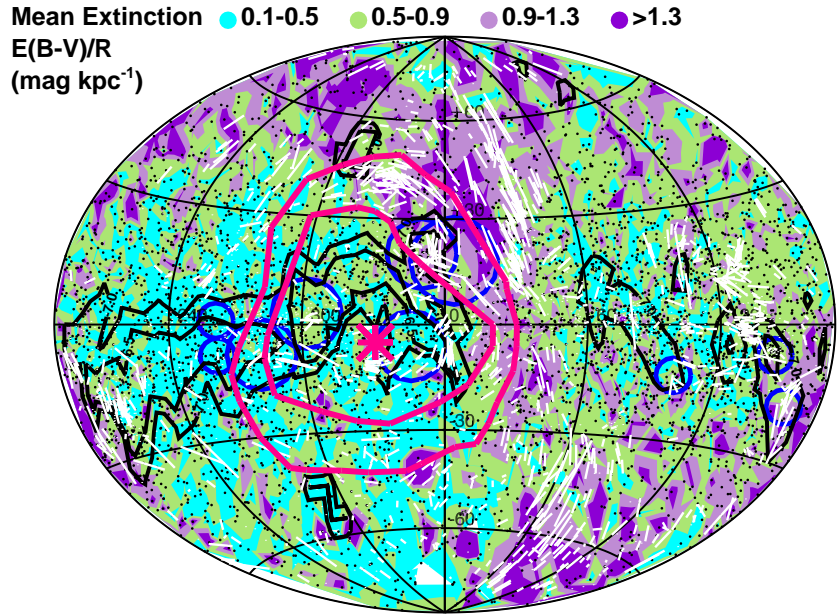}}
\caption{Mean extinction, \ebv/$R$, for stars with $R < 0.5$, where $R$ 
is the distance in kpc.
Black contours show the integrated stellar radiation at 1565 A measured
by the TD-1 satellite (Gondhalekar et al., 1980).  Contours indicate 
flux levels of $10^{-7}$ and $10^{-6.5}$ ergs \cmtwo\ s$^{-1}$ A$^{-1}$ 
sterad$^{-1}$.  Polarization data for stars within 500 pc of the Sun 
are plotted as white bars (Heiles 2000).
Pink contours show the ring that may be formed by the 
Loop I supernova remnant interaction with the Local Bubble (Egger
and Aschenbach, 1995).   The pink asterisk indicates the
LSR upwind direction of the Local Fluff (see text).
Blue circles show stellar OB associations
within 500 pc of the Sun.  An Aitoff projection is used,
with the galactic center at the center of the plot, and \glong\ 
increasing to the right.  Note that the ISM density is very low out to 500 pc
in the interval  $\ell = 180^\circ \rightarrow
300^\circ$, particularly towards low galactic latitudes.  This
plot is based on photometric data in the Hipparcos catalog.  The 
stars (small black dots) are preselected for photometric quality, and color 
excess values are smoothed for stars with overlapping distance uncertainties and
within 13$^\circ$ of each other on the sky.  
The average distances of stars sampling latitudes in the intervals 
$| b | < [30^\circ,~30^\circ - 60^\circ, ~> 60^\circ]$ are [210, ~164,~138] pc, so that the ISM in high latitude sightlines tends
to be closer.  Uncertainties on \ebv\ for the plotted stars are 
typically $<0.02$ mag because stars are also required to have 
photometric distances that are consistent with astrometric distances 
to within $\sim30$\%.
Intrinsic stellar colors are from Cox (2000).}

\label{fig:lb2}
\end{figure}

Since the formation of Gould's Belt, the
Sun has traveled hundreds of parsecs through the LSR, and the LSR has
completed $\sim 10 - 25$\% of its orbit around the galactic center.
Molecular clouds disrupted by the initial blast wave now rim Gould's
Belt.  Epochs of star formation in the Scorpius-Centaurus Association
during the past 1--15 Myrs further evacuated the Local Bubble
void, and displaced ISM from the SCA into giant nested \HI\ shells
(de Geus, 1992). One of these shells, Loop I (the North Polar Spur), was formed by a recent supernova ($< 1$ Myrs
ago) and is an intense source of polarized synchrotron and soft X-ray
(SXR) emission.  A ring-like shadow, caused by foreground ISM, is seen
in the Loop I SXR emission.  The origin of this ring has been
suggested to be the result of Loop I merging with the separate Local
Bubble (e.g. Egger and Aschenbach, 1995).  The LSR
upwind direction of the Local Fluff
is at the center of this ring, and polarization 
data show that the ring is a magnetic loop (Fig. \ref{fig:lb2}).
\footnote{In accordance with general practice, here I use the 
Standard solar apex motion (velocity 19.7 \kms, towards
 \glong=57$^\circ$ and \glat=22 $^\circ$) 
to correct heliocentric velocities to the LSR.
This gives a Local Fluff LSR 
upwind direction towards \glong, \glat $ = 331^\circ, ~ -5^\circ$, and LSR 
velocity --19.4 \kms.  In Frisch et al. (2002), we instead used an
LSR conversion based on Hipparcos data, and found an LSR upwind
direction \glong=2.3$^\circ$, \glat=--5.2$^\circ$ (velocity
--17 \kms), which is
directed towards the ISM shadow itself.  This difference in the two
possible Local Fluff upwind directions reveals an obvious flaw in 
comparing a dynamically defined direction with
a statically defined diffuse object such as the more distant SXRB shadow.}

\subsection{Boundaries}

The locations of the Local Bubble boundaries have been diagnosed with
a range of different ISM markers, including color excess (Lucke,
1978), ultraviolet observations of interstellar
\HI\ lines in hot stars (York \& Frisch, 1983, Paresce,
1984), radio \HI\ 21-cm and
optical \NaI\ data (Vergely et al., 2001),
extreme ultraviolet (EUV, $\lambda < 912$ A) emission of white dwarf
and M-stars (Warwick, 1993), measurements of
polarization of starlight (Leroy, 1999), and the
trace ionization species \NaI\ (Sfeir et al., 1999, Lallement et
al. 2003).  These studies
differ in sampling densities and spatial smoothing methods.  Each
marker is an imprecise tracer of the total ISM mass
density, since the ISM is highly inhomogeneous
over the scale lengths of the Local Bubble and small scale structure
is poorly understood.
I focus here on the reddening data.


An accessible measure of starlight reddening is color excess,
\ebv, which measures the differential extinction of starlight in the
blue versus visual bands, and is sensitive to
interstellar dust grains (ISDG) of radii $a \sim 0.20 $ \micron.
Interstellar gas and dust are generally well mixed, so that the 
threshold reddening for the Local Bubble
walls found from \ebv\ data is consistent
with the locations found from gas markers.  The exception is that
dust is found in both neutral and ionized regions, while the
commonly available gas markers (\HI, \NaI) are weighted towards
neutral regions.  

Grains and gas are well mixed partly because both populations couple
to the interstellar magnetic field (\Bis) in cold and warm clouds.  In
cold clouds with density \nH=100 \cc\ and temperature $T \sim 100$ K,
the $a \sim 0.2$ \micron\ dust grain with density 2 g \cc\ will sweep
up its own mass in gas in $\sim 0.08$ Myrs.  If the same grain has
charge Z=20, the gyrofrequency for a magnetic field of strength B=2.5
\mG\ is $\sim 1/3300$ yrs.  For a warm neutral cloud (\nH$\sim 0.25$
\cc, T$\sim 6300$ K), the grain accumulates its own mass in $\sim 4$
Myrs.  In both cloud types, grains couple to \Bis.  Gas also couples
to \Bis, since elastic collisions couple neutrals and ions over
time-scales of years, and minimum ionization levels of $\sim 10^{-4}$
bind gas to \Bis\ (Spitzer, 1978). 

In Fig. \ref{fig:lb2}, the reddening per unit distance, \ebv/$R$, where $R$
is the star distance in kpc, is
shown on an Aitoff projection for O, B, and A stars within 500 pc.
\ebv\ values are based on Hipparcos photometric data (Perryman et al., 1997). 
The lowest mean ISM densities in the galactic plane
are between longitudes of $210 ^\circ$ and $360^\circ$
are evident.  Star groups (blue circles) in the low density sightlines
include Sco OB2, Vela OB2, and Trumpler 10 (de Zeeuw et al.,
1999). The lowest mean densities in this data
set, outside of the Local Fluff, correspond to $0.006$ atoms \cc.  At
the galactic poles, $ |b| > 75^\circ$, the edges of the Local Bubble,
where \ebv$ > 0.05$ mag (or approximately log \NH$> 20.50$
\cmtwo),\footnote{The ratio \ebv/\NHI\ varies in sightlines with low
mean extinctions because of variations in both mean grain sizes and
hydrogen ionization.} are at 80--95 pc towards both the north and
south poles.

\begin{figure}[ht!]
\centerline{\includegraphics[angle=0,clip=,width=0.79\textwidth]{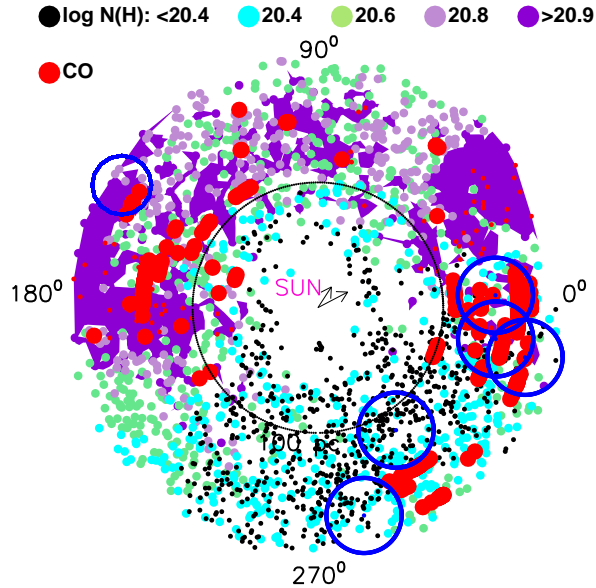}}
\caption{The distribution of ISM surrounding the Local
Bubble void, based on stars within 200 pc and within $45^\circ$ of the galactic
plane.  Molecular clouds of CO and H$_2$ are plotted as red
symbols (Dame et al. 2001).  The colored dots show cumulative hydrogen
column densities towards O, B, and A stars based on
\ebv\ (Fig. \ref{fig:lb2}) and the gas-to-dust ratio measured by
the $Copernicus$ satellite for stars with low mean
extinctions (see text).  The mean gas-to-dust relation overestimates
\NH\ at low column densities (\NH$< 10^{20}$ \cmtwo, Bohlin et al. 1978).  
Sightlines with \NH$ > 10^{20.9}$ \cmtwo\ have been plotted with filled purple
contours.  The arrows show two different values for the direction of
the Sun's motion through the LSR, with the longer
arrow ($v = 19.5$ \kms) indicating the Standard solar apex motion.  
Blue circles indicate OB associations within 200 pc of the Sun.  
The black circle indicates 100 pc.}\label{fig:lb1}
\end{figure}

The LB boundaries in the galactic plane are shown in Fig. \ref{fig:lb1}, 
for $\sim 2000$ O, B, and A stars within$\sim 200$ pc and $45 ^\circ$ of 
the galactic plane, using a threshold cumulative value of log \NH$ >  20.4$ \cmtwo\
corresponding to \ebv$ > 0.04 $ mag when \NH/\ebv$ \sim 5.8 \times 10^{21} $
  \cmtwo\ mag$^{-1}$ K.  This gas-to-dust ratio is good
to within factors of $\sim 2$ for
\ebv$>0.1$ mag and $\sim 3$ for \ebv$< 0.1$ mag (Bohlin et al., 1978).  
Note the well known deficiency of ISM out to distances beyond 200 pc 
in the third and parts of the fourth galactic quadrants
(Frisch and York, 1983).  For cloudy
sightlines (high mean \ebv\ values), the fraction of the H atoms in
\HH\ ($f_\mathrm{H_2}$) rises above $\sim 1$\% at \ebv$\sim 0.1$ mag.  The
classic term ``intercloud'' refers to low column density sightlines with
relatively little \HH\ ($f_\mathrm{H_2} < 1$\%).  
Molecular clouds of CO and \HH\
are also shown, and are plotted as filled red circles (Dame et al.,
2001). Well-known molecular clouds at the rim of
the Local Bubble include dust in Scorpius (\glong$\sim 350^\circ$,
d$\sim 120 $ pc), Taurus (\glong$\sim 160^\circ$, d$\sim 120$ pc), and
Chameleon (\glong$\sim 305 ^\circ$, \glat$ \sim -15^\circ$, d$\sim
165$ pc).

The mean value of \ebv/(\NHI+2\NHH) varies by $ \sim 15$\% between sightlines
with low and high fractions of \HH\ (Bohlin et al., 1978), because
of variations in the mean grain size and radiation field.  The 
$\lambda \sim 1565$ A radiation field depends on location with respect 
to the Local Bubble walls (\S \ref{sec:rf}), and the 912-1108 A radiation
field capable of photodissociating \HH\ should behave in a
similar fashion.

\section{Loop I and the Local Magnetic Field }

\subsection{Loop I and the High-Latitude Limits of the Local Bubble}

Above the galactic plane in the galactic-center hemisphere, \glat$>
20^\circ$, the LB walls are established by neutral gas of the Loop I
superbubble.  The interval \glong$ \sim 270^\circ \rightarrow
50^\circ$ is encircled by high-latitude nested shells of gas and dust.  
Loop I is $\sim 80 ^\circ$ in radius and centered 120 pc away at
\glong$=320^\circ$, \glat$=5^\circ$ for the neutral gas (Berkhuijsen
1971; Heiles, 1998; de Geus, 1992).  The central
regions of these evacuated shells are deficient in ISM, creating the extension
of the Local Bubble towards \glong$\sim 340^\circ$
(Fig. \ref{fig:lb1}).

\subsection{Magnetic Field}

Loop I dominates the magnetic field structure near the Sun, and is a
source of intense radio continuum and soft X-ray emission.  
The Loop I magnetic field, comprised of components parallel (\Bpar)
and perpendicular (\Bperp) to the sightline, is traced by polarized synchrotron emission, starlight
polarization caused by magnetically aligned dust grains, Faraday
rotation, and Zeeman splitting of the \HI\ 21-cm line.
Fig. \ref{fig:lb2} shows the starlight polarization vectors (from
Heiles, 2000).  Magnetically aligned interstellar dust grains (ISDGs) are birefringent at optical
wavelengths, with lower opacities found for the polarization component
parallel to \Bis.  The Loop I magnetic field direction is shown by the
gradient in the rotation angle of the optical polarization vectors,
which follows the interaction ring feature.  Comparisons between
the optical polarization data (tracing \Bpar) and synchrotron emission 
(tracing \Bperp), indicate that \Bis\ is nearly
in the plane of the sky in Loop I (Berkhuijsen, 1972, Heiles \&
Crutcher, 2005).  

The closest measured \Bis\ strengths are towards Loop I.
Heiles et al. (1980) found a volume averaged
field strength of \Bis$ \sim 4$ \mG\ in a tangential direction through
the shell (extending $\sim 70 \pm 30$ pc towards \glong$= 34^\circ$,
\glat$=42^\circ$).  Faraday rotations of extragalactic radio sources
indicate that \Bpar\ is small, with an average value of $
B_\mathrm{||} = 0.9 \pm 0.3 $ \mG\ from rotation measure data (Frick
et al., 2001).  Magnetic pressure dominates in the neutral shell gas.
In the ionized gas producing the
radio continuum emission, the magnetic, gas, and cosmic ray pressures
are all significant.  Loop I is a decelerated shock generated by
sequential epochs of star formation in SCA (de Geus, 1992)

\subsection{The LIC and the Magnetic Field at the Sun}

The LIC is very low density, \nH $\sim 0.25 - 0.30$ \cc.  Magnetic
fields in high density ISM show evidence of flux freezing; however
\Bis\ in low density ISM appears uncorrelated with density (Heiles \&
Crutcher, 2005).  Pulsar dispersion measures indicate that the uniform
component of the magnetic field near the Sun is \Bis$ \sim 1.4$ \mG,
with correlation lengths of $\sim 100$ pc (Rand \& Kulkarni,
1989). In general, structure functions
created from data on radio continuum polarization near 21 cm show that
magneto-ionized structures in interarm sightlines must be very large
(e.g. $\sim 100$ pc, Haverkorn et al., 2006).
This would indicate that the uniform \Bis\ component is appropriate
for the low density (similar to interarm) region around the Sun.

The physical conditions of the LIC have been modeled by developing a
series of radiative transfer models that are constrained by
observations of \HeI\ and pickup ion and anomalous cosmic ray data
inside of the heliosphere, and observations of the LIC towards
$\epsilon$ CMa.  These models are discussed in detail in Slavin and
Frisch (this volume, and 2007, hereafter SF07a,b).
The best of these models give \nHI$=0.19
- 0.20$ \cc, \nel$=0.07 \pm 0.02$ \cc, \nHeI=0.015 \cc, for cloud
temperatures $\sim$6300 K.  If the magnetic and gas pressures are
equal in the LIC, then the LIC field strength is $B_\mathrm{LIC} \sim
2.8$ \mG.  This value is also consistent with the interface magnetic field
strength of 2.5 \mG, adopted in the best model (model 26).  However,
it is somewhat above the strength of the uniform component of \Bis.
Since the ISM flow past the Sun has an origin associated with the
breakaway of a parcel of ISM from the Loop I magnetic superbubble 
(Frisch, 1981),
perhaps \Bis\ at the Sun is stronger and perturbed compared to the
uniform field, but at lower pressure than the confined parts of the
Loop I bubble.

\begin{figure}[b!]
\begin{center}
\includegraphics[height=3.2in,width=3.2in]{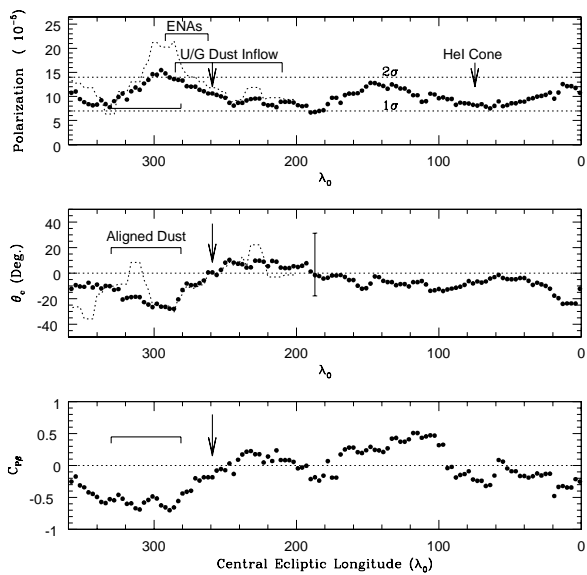}
\end{center}
\caption{Interstellar polarization towards nearby stars
(data from Tinbergen, 1982) compared to the ecliptic position
of the star.  A systematic enhancement of the polarization strength 
is found close to the ecliptic at an offset in ecliptic longitude
$\lambda$ of $\sim +35^\circ$ compared 
to the inflowing upwind gas and large dust grain directions.  
Top: The average polarization $P$ for stars with $ |\beta|
< 50^\circ$ is plotted as dots, and for stars with $| \beta| < 20^\circ$ 
as a dashed line.  Data are averaged over $\pm 20^\circ$ around
the central  ecliptic
longitude, $\lambda_\circ$.  The direction of maximum $P$ is shifted by
$\sim 25 -30^\circ$ from the upwind direction of the large interstellar
dust grains detected by Ulysses/Galileo (Frisch et al., 1999). The
upwind gas and large-grain directions differ by $< 5^\circ$.  Middle: The
averaged polarization position angle in celestial coordinates,
$\Theta_\mathrm{c}$.  In the region of maximum polarization, $\lambda \sim
280^\circ \rightarrow 310^\circ$, the grains show consistent position
angles.  Bottom: The correlation coefficient between $P$ (top) and
$\beta$ is shown as a function of the ecliptic longitude.  The
strongest polarization is found at negative ecliptic latitudes.  
For more details see Frisch (2005, 2006).} \label{fig:f3} 
\end{figure}

Very weak interstellar polarization caused by magnetically aligned
dust grains has been observed towards stars within $\sim 35$ pc
(Tinbergen, 1982).  The polarization was
originally understood to arise in the Local Fluff, since the
polarization region coincides with the upwind direction of the flow
where column densities are highest.  More recently, the polarization
properties were found to have a systematic relation to ecliptic
geometry.  The region of maximum polarization is found to be located
at ecliptic longitudes that are offset by $\sim +35^\circ$ from the
large dust grains flowing into the heliosphere,
and from the gas upwind direction (Fig. \ref{fig:f3}, Frisch et al., 1999).
Stars with high polarizations also show consistent polarization
angles, and in general polarization is higher for negative
ecliptic latitudes.
These polarization data are consistent with the interpretation that
polarizing grains are trapped in \Bis\ as it drapes over the
heliosphere (Frisch, 2005, 2006).
When magnetically prealigned (by \Bis) grains approach the heliosphere,
the gas densities are too low to collisionally disrupt the alignment,
and polarization should indicate the direction of
\Bis\ at the heliosphere.  If the alignment mechanism is sufficiently
rapid, the alignment strength and direction will also adjust to the interstellar
magnetic field direction as it drapes over the heliosphere.  Although
this interpretation of the polarization data is not confirmed, it fits
the physics of dust grains interacting with the heliosphere.  Small
charged grains such as those that polarize starlight ($a < 0.2$
\micron) couple to \Bis\ and are excluded from the heliosphere, while
large grains enter the heliosphere where they are measured by various
spacecraft (Krueger, this volume).  The characteristics of such
polarization may vary with solar cycle phase.


\section{Radiation Environment of the Local Bubble}

\subsection{Inhomogeneous Radiation field and Local Fluff Ionization}\label{sec:rf}

The interstellar radiation field (ISRF) is key to 
understanding the physical properties of the LIC and Local Fluff.  The
sources of the ISRF at the Sun include plasma emission from the Local
Bubble interior and supernova remnants, stellar radiation, including
from hot white dwarf stars, and emission from a conductive interface
between the local fluff and the hot plasma.  The spectrum of this
field at the surface of the LIC is shown in Slavin and Frisch (this
volume).

The spectrum of the ISRF is inhomogeneous because of the
energy-dependent opacity of the ISM.  For instance, radiation with
$\lambda < 912$ A (584 A) determines the ionizations of H (He).
Energetic photons capable of ionizing \HI\ (\HeI) require \NHI$\sim
17.2$ (17.7) \cmtwo\ to reach an opacity $\tau \sim 1$.  The
dependence of $\tau_\mathrm{912 A} / \tau_\mathrm{504 A} $ on \NHI\
drives the need for LIC photoionization models to determine the
heliosphere boundary conditions.  Stars within $\sim 10$ pc (e.g. Wood
et al. 2005) show local column density
variations of \mbox{log \NHI$ \sim 17.07 - 18.22$} \cmtwo\ dex
(assuming log D/H$ =-4.7$).  This yields a range locally of $\tau_{912
\mathrm{A}} = 0.7 - 10.5$, and shows that ionization must vary between
the individual cloudlets comprising the Local Fluff.  This variation
is confirmed by \NII\ data, which are excellent tracers of \HII\
through charge-exchange.  Stars within 70 pc show \NII/\NI$\sim 0.1 -
2$ (SF07a,b). Our LIC radiative transfer models indicate that in the
LIC \HI\ ionization provides $\sim 66$\% of the cloud heating, and the LIC is
$\sim 20 - 30$\% ionized (SF07a,b).

Another example of the inhomogeneous ISRF is provided by the photon
flux at $\lambda \sim 1565$ A, which TD-1 satellite data show depends
on position in the Local Bubble (Gondhalekar et al.,
1980). The ISRF at 1565 A is
dominated by hot stars, B or earlier.  Radiation at $\lambda \leq
1620$ A regulates the photoionization rate of interstellar \MgI, and
is an important parameter for the \MgII/\MgI\ diagnostic of the
interstellar electron density.  The ISDG albedo at $\lambda \sim 1565$
A is $\sim 0.5$.  Figure \ref{fig:lb2} shows the flux of 1565 A photons
at the Sun, plotted as black contours. 
The brightest regions of the sky at 1565 A are in the
third and fourth galactic quadrants, \glong$ \sim 180^\circ -
360^\circ$, where the mean extinction in the interior of the Local
Bubble is low, \ebv$/D <0.4$ mag kpc$^{-1}$.

The fact that the 1565 A radiation field is enhanced near the
galactic plane for $\ell \sim 180 \rightarrow 360^\circ$
is relevant to our understanding of the more energetic photons
associated with the soft X-ray background (SXRB).  Isolated bright
SXRB regions are seen, such as the Orion-Eridanus and Loop I
enhancements, however no regional enhancement in the SXRB flux is seen
corresponding to the bright $\lambda 1565$ A regions.

\subsection{Diffuse Soft X-Ray Background}

The diffuse soft X-ray background (SXRB) is significant
both as an ionizing and heating radiation field, and as a
fossil that traces the supernovae that formed the Local Bubble.

The spectrum of the soft X-ray background (SXRB) emission determined using
broadband sounding rocket observations at energies $\sim 0.08 - 0.2$
keV revealed an excess of count rates at low energies in the galactic
plane, compared with the spectrum predicted by an absorbed $T \sim 10^6$ K
plasma.  This effect was interpreted as indicating a local X-ray
plasma with less than $\sim 7 \times 10^{18}$ \cmtwo\ of foreground
hydrogen (Juda et al., 1991).  The complex SXRB spectrum includes
this local hot plasma, inferred to fill the LB, along with
contributions from absorbed galactic halo emission, extragalactic 
emission, and absorbed local supernova remnants such as Loop I which 
dominates the northern hemisphere sky.

Fast-forwarding to the present, the LB contributions to the SXRB have
been found from broadband ROSAT data (0.18 -- 0.3 keV) and XMM Newton
spectra of dark clouds that help isolate the LB flux from foreground
and background fluxes, combined with SXRB models that include
foreground contamination from charge-exchange between the solar wind
and neutral interstellar or geocoronal gas (e.g. Robertson and
Cravens, 2003; Snowden et al., 2004; Bellm and Vaillancourt, 2005;
Henley et al. 2007; Lallement, this volume ).  The contribution of CEX
to the SXRB is significant above $\sim 0.3$ keV, and is limited by the
SXRB flux that does not anticorrelate with \NHI.  Models of the
contributions of heliospheric charge exchange to the SXRB measured by
ROSAT indicate that $\sim 50- 65$\% of the SXRB in the galactic plane
and $\sim 25$\% at high galactic latitudes may arise from CEX,
depending on the adopted model for CXE fluxes, the solar cycle phase,
and ecliptic latitude of the look direction.  LB plasma models with
depleted abundances predict approximate consistency between the ROSAT
and sounding rocket SXRB data.  Comparisons between XMM Newton spectra
acquired while pointing on and off of a dark cloud in principle allow
the removal of foreground heliospheric CEX.  For a solar abundance
plasma model, based on recent values (Asplund, this volume), the
radius of the cavity filled with plasma is $\sim 100$ pc, with plasma
density \nel=0.013 \cc, pressure $2.9 \times 10^4$ \cc\ K, a cooling
time of 17 Myrs, and a sound-crossing time of 1.2 Myrs.  The LB plasma
properties are a topic of ongoing research because the foreground
contamination from CEX and contributions to the SXRB spectra
contributed by the halo and disk gas are poorly understood.

The LIC ram pressure, $P_\mathrm{ram}$, may be important for the pressure
 equilibrium between the LIC and LB plasma.  At the Sun, the SF07 
LIC models find a total density of $n$(\HI+\HII)$ \sim 0.26$ atoms \cc, 
and \nel$\sim
  0.07$ \cc, for $T = 6300$ K, giving a static LIC pressure of
  $\sim 2200$ \cc\ K (including He).  For a relative LIC-LB velocity of $\sim
  20.7$ \kms, $P_\mathrm{ram} \sim 20,000$ \cc\ K,
so that $P_\mathrm{ram}$ helps offset the high LB thermal pressure in 
one dimension.
However, for the LIC ram pressure to remove the
  longstanding mystery about the pressure equilibrium between the LIC
  and LB plasma, additional pressure contributions 
(e.g. from magnetic and cosmic ray)
  are required for directions perpendicular to the relative
  LIC-hot gas velocity vector.


\section{The LIC and the ISM Flow Past Sun}  

\subsection{Kinematics of Local Fluff}

The LIC is part of an localized ISM flow that has been denoted the Local Fluff,
or Cluster of Local Interstellar Clouds (CLIC).  The best fitting heliocentric 
flow vector for ISM within $\sim 30$ pc is $-28.1 \pm 4.6$ \kms, from the
direction \glong$ =12.4 ^\circ$, \glat$ = 11.6^\circ$ (Frisch et al. 2002). 
In the LSR, the upwind direction
is towards the center of Loop I (footnote 2, \S \ref{sec:origin}).  
The flow is decelerating; in the rest frame of the flow velocity, the
fastest components in the upwind and downwind directions are
blue-shifted by over 10 \kms.  Individual cloudlets contribute to the
flow, including the LIC, the ``G-cloud'' within 1.3 pc in
the upwind direction, and the Apex cloud within 5 pc 
in the upwind direction and extending towards $\glong \sim
30^\circ$.  These cloudlets have the same upwind
directions to within $\sim \pm 10^\circ$, indicating a common origin
for the cloudlets comprising the Local Fluff.  Alternate
interpretations using Local Fluff kinematics and temperatures 
have parsed the flow into $\sim 15$ spatially distorted components
(Linsky, this volume).  However, velocity components
towards stars in the sidewind direction can not clearly distinguish
between individual clouds because of velocity blending.

\subsection{Interstellar Abundances}

The abundance pattern of elements in interstellar gas is characterized
by abundances that decrease with increases in the mean gas density
(\nHmean) or elemental condensation temperature ($T_\mathrm{cond}$).
Most depletion studies are based on long sightlines with blended
velocity components.  Our LIC radiative transfer models derive
elemental abundances corrected for ionization, for a single low
density cloud in space (SF07a,b).  For the best models in SF07b, LIC
abundances are O/H=295-437 ppm, compared to solar abundances 460 ppm
(Asplund, this volume), and N/H=40--66 ppm, compared to solar values
61 ppm.  The LIC S/H ratios are 14--22 ppm, compared to solar values
14 ppm. If the LIC has a solar composition, as indicated by anomalous
cosmic ray data for $^{22}$Ne and $^{18}$O (Cummings, this volume),
then Asplund solar abundances are preferred over earlier values.
Carbon is found to be overabundant by a factor of $\sim 2.6$ compared
to solar abundances, which helps maintain the LIC cloud temperature near the
observed temperature of 6300 K through C fine-structure cooling.  The
C-abundance anomaly appears to be due to the destruction of
carbonaceous grains by interstellar shocks.  The carbon overabundance
is consistent with the deficit of small carbonaceous grains causing
the 2200 A bump and far-ultraviolet rise in the ultraviolet extinction
curves in some regions.


{\bf Acknowledgements:  }I thank the organizers of the Geiss-fest
for a stimulating meeting, and ISSI for recognizing long ago that
the interaction of local ISM and the heliosphere is a fascinating
topic.  NASA grants NNG05GD36G, NNG06GE33G, and 
NAG5-13107 supported this work.  The Hipparcos data-parsing
tool was developed by Prof. Philip Chi-Wing Fu, under the 
auspices of NASA grant NAG5-11999.

\end{article}

\begin{thebibliography}{40}
\expandafter\ifx\csname natexlab\endcsname\relax\def\natexlab#1{#1}\fi

\bibitem[{{Bellm} \& {Vaillancourt}(2005)}]{BellmVaillancourt:2005}
{Bellm}, E.~C. \& {Vaillancourt}, J.~E. 2005, \apj, 622, 959

\bibitem[{{Bohlin} {et~al.}(1978){Bohlin}, {Savage}, \&
  {Drake}}]{BohlinSavageDrake:1978}
{Bohlin}, R.~C., {Savage}, B.~D., \& {Drake}, J.~F. 1978, \apj, 224, 132

\bibitem[{{Breitschwerdt} {et~al.}(2000){Breitschwerdt}, {Freyberg}, \&
  {Egger}}]{Breitschwerdtetal:2000}
{Breitschwerdt}, D., {Freyberg}, M.~J., \& {Egger}, R. 2000, \aap, 361, 303

\bibitem[{{Dame} {et~al.}(2001){Dame}, {Hartmann}, \&
  {Thaddeus}}]{Dameetal:2001}
{Dame}, T.~M., {Hartmann}, D., \& {Thaddeus}, P. 2001, \apj, 547, 792

\bibitem[{{de Geus}(1992)}]{deGeus:1992}
{de Geus}, E.~J. 1992, \aap, 262, 258

\bibitem[{{de Zeeuw} {et~al.}(1999){de Zeeuw}, {Hoogerwerf}, {de Bruijne},
  {Brown}, \& {Blaauw}}]{deZeeuw:1999}
{de Zeeuw}, P.~T., {Hoogerwerf}, R., {de Bruijne}, J. H.~J., {Brown}, A. G.~A.,
  \& {Blaauw}, A. 1999, \aj, 117, 354

\bibitem[{{Egger} \& {Aschenbach}(1995)}]{Egger:1995}
{Egger}, R.~J. \& {Aschenbach}, B. 1995, \aap, 294, L25

\bibitem[{{Fitzgerald}(1968)}]{Fitzgerald:1968}
{Fitzgerald}, M.~P. 1968, \aj, 73, 983

\bibitem[{{Frick} {et~al.}(2001){Frick}, {Stepanov}, {Shukurov}, \&
  {Sokoloff}}]{Fricketal:2001}
{Frick}, P., {Stepanov}, R., {Shukurov}, A., \& {Sokoloff}, D. 2001, \mnras,
  325, 649

\bibitem[{{Frisch}(1981)}]{Frisch:1981}
{Frisch}, P.~C. 1981, Nature, 293, 377

\bibitem[{{Frisch}(1995)}]{Frisch:1995}
{Frisch}, P.~C. 1995, \ssr, 72, 499

\bibitem[{{Frisch}(2005)}]{Frisch:2005L}
{Frisch}, P.~C. 2005, \apjl, 632, L143

\bibitem[{{Frisch}(2006)}]{Frisch:2006II}
{Frisch}, P.~C. 2006, \apj, submitted

\bibitem[{{Frisch} {et~al.}(1999){Frisch}, {Dorschner}, {Geiss}, {Greenberg},
  {Gr\"un}, {Landgraf}, {Hoppe}, {Jones}, {Kr{\"{a}}tschmer}, {Linde},
  {Morfill}, {Reach}, {Slavin}, {Svestka}, {Witt}, \& {Zank}}]{Frischetal:1999}
{Frisch}, P.~C., {Dorschner}, J.~M., {Geiss}, J., {et~al.} 1999, \apj, 525, 492

\bibitem[{{Frisch} {et~al.}(2002){Frisch}, {Grodnicki}, \& {Welty}}]{FGW:2002}
{Frisch}, P.~C., {Grodnicki}, L., \& {Welty}, D.~E. 2002, \apj, 574, 834

\bibitem[{{Frisch} \& {York}(1983)}]{FrischYork:1983}
{Frisch}, P.~C. \& {York}, D.~G. 1983, \apjl, 271, L59

\bibitem[{{Gondhalekar} {et~al.}(1980){Gondhalekar}, {Phillips}, \&
  {Wilson}}]{Gondhalekaretal:1980}
{Gondhalekar}, P.~M., {Phillips}, A.~P., \& {Wilson}, R. 1980, \aap, 85, 272

\bibitem[{{Grenier}(2004)}]{Grenier:2004}
{Grenier}, I.~A. 2004, ArXiv Astrophysics e-prints

\bibitem[{{Haverkorn} {et~al.}(2006){Haverkorn}, {Gaensler}, {Brown},
  {Bizunok}, {McClure-Griffiths}, {Dickey}, \& {Green}}]{Haverkorn:2006}
{Haverkorn}, M., {Gaensler}, B.~M., {Brown}, J.~C., {et~al.} 2006, \apjl, 637,
  L33

\bibitem[{{Heiles}(1998)}]{Heiles:1998}
{Heiles}, C. 1998, \apj, 498, 689

\bibitem[{{Heiles}(2000)}]{Heiles:2000pol}
{Heiles}, C. 2000, \aj, 119, 923

\bibitem[{{Heiles} {et~al.}(1980){Heiles}, {Chu}, {Troland}, {Reynolds}, \&
  {Yegingil}}]{Heilesetal:1980}
{Heiles}, C., {Chu}, Y.~., {Troland}, T.~H., {Reynolds}, R.~J., \& {Yegingil},
  I. 1980, \apj, 242, 533

\bibitem[{{Heiles} \& {Crutcher}(2005)}]{HeilesCrutcher:2005}
{Heiles}, C. \& {Crutcher}, R. 2005, ArXiv Astrophysics e-prints

\bibitem[{{Henley} {et~al.}(2007){Henley}, {Shelton}, \&
  {Kuntz}}]{HenleyShelton:2007}
{Henley}, D.~B., {Shelton}, R.~L., \& {Kuntz}, K.~D. 2007, ArXiv Astrophysics
  e-prints

\bibitem[{{Juda} {et~al.}(1991){Juda}, {Bloch}, {Edwards}, {McCammon},
  {Sanders}, {Snowden}, \& {Zhang}}]{JudaMcCammon:1991}
{Juda}, M., {Bloch}, J.~J., {Edwards}, B.~C., {et~al.} 1991, \apj, 367, 182

\bibitem[{{Lallement} {et~al.}(2003){Lallement}, {Welsh}, {Vergely}, {Crifo},
  \& {Sfeir}}]{Lallementetal:2003}
{Lallement}, R., {Welsh}, B.~Y., {Vergely}, J.~L., {Crifo}, F., \& {Sfeir}, D.
  2003, \aap, 411, 447

\bibitem[{{Leroy}(1999)}]{Leroy:1999}
{Leroy}, J.~L. 1999, \aap, 346, 955

\bibitem[{{Lucke}(1978)}]{Lucke:1978}
{Lucke}, P.~B. 1978, \aap, 64, 367

\bibitem[{{Ma{\'{\i}}z-Apell{\'a}niz}(2001)}]{MaizApel:2001}
{Ma{\'{\i}}z-Apell{\'a}niz}, J. 2001, \apjl, 560, L83

\bibitem[{{Paresce}(1984)}]{Paresce:1984}
{Paresce}, F. 1984, \aj, 89, 1022

\bibitem[{{Perryman}(1997)}]{Perrymanetal:1997}
{Perryman}, M.~A.~C. 1997, \aap, 323, L49

\bibitem[{{Rand} \& {Kulkarni}(1989)}]{RandKulkarni:1989}
{Rand}, R.~J. \& {Kulkarni}, S.~R. 1989, \apj, 343, 760

\bibitem[{{Robertson} \& {Cravens}(2003)}]{RobertsonCravens:2003}
{Robertson}, I.~P. \& {Cravens}, T.~E. 2003, Journal of Geophysical Research
  (Space Physics), 108, 6

\bibitem[{{Sfeir} {et~al.}(1999){Sfeir}, {Lallement}, {Crifo}, \&
  {Welsh}}]{Sfeiretal:1999}
{Sfeir}, D.~M., {Lallement}, R., {Crifo}, F., \& {Welsh}, B.~Y. 1999, \aap,
  346, 785

\bibitem[{{Slavin} \& {Frisch}(2006)}]{SlavinFrisch:2007}
{Slavin}, J.~D. \& {Frisch}, P.~C. 2007, \aap, in preparation

\bibitem[{{Snowden} {et~al.}(2004){Snowden}, {Collier}, \&
  {Kuntz}}]{SnowdenCollier:2004}
{Snowden}, S.~L., {Collier}, M.~R., \& {Kuntz}, K.~D. 2004, \apj, 610, 1182

\bibitem[{Spitzer(1978)}]{Spitzer:1978}
Spitzer, L. 1978, Physical Processes in the Interstellar Medium (John
  Wiley \& Sons, Inc.)

\bibitem[{Tinbergen(1982)}]{Tinbergen:1982}
Tinbergen, J. 1982, \aap, 105, 53

\bibitem[{{Vergely} {et~al.}(2001){Vergely}, {Freire Ferrero}, {Siebert}, \&
  {Valette}}]{Vergelyetal:2001}
{Vergely}, J.-L., {Freire Ferrero}, R., {Siebert}, A., \& {Valette}, B. 2001,
  \aap, 366, 1016

\bibitem[{{Warwick} {et~al.}(1993){Warwick}, {Barber}, {Hodgkin}, \&
  {Pye}}]{Warwicketal:1993}
{Warwick}, R.~S., {Barber}, C.~R., {Hodgkin}, S.~T., \& {Pye}, J.~P. 1993,
  \mnras, 262, 289

\bibitem[{{Wood} {et~al.}(2005){Wood}, {Redfield}, {Linsky}, {M{\"u}ller}, \&
  {Zank}}]{Woodetal:2005}
{Wood}, B.~E., {Redfield}, S., {Linsky}, J.~L., {M{\"u}ller}, H.-R., \& {Zank},
  G.~P. 2005, \apjs, 159, 118

\end{thebibliography}
\end{document}